\documentclass[conference]{IEEEtran}
\IEEEoverridecommandlockouts

\usepackage{cite}
\usepackage{amsmath,amssymb,amsfonts}
\usepackage[nolist]{acronym}
\usepackage{algorithmic}
\usepackage{enumitem} 

\usepackage{graphicx}  
\usepackage{subfig}    
\usepackage{subcaption} 

\usepackage{subfig}
\usepackage{graphicx}
\usepackage{multicol}  
\usepackage{epsfig}
\usepackage{textcomp}
\usepackage{xcolor}
\usepackage{booktabs}
\usepackage{geometry}
\def\BibTeX{{\rm B\kern-.05em{\sc i\kern-.025em b}\kern-.08em
    T\kern-.1667em\lower.7ex\hbox{E}\kern-.125emX}}

\begin{document}

\title{Interpretable Attention-Based Multi-Agent PPO for Latency Spike Resolution in 6G RAN Slicing
\thanks{\textcolor{black}{The work is funded by the UK Engineering and Physical Sciences Research Council (EPSRC) under grants EP/Y037421/1, EP/Y036514/1, EP/X040518/1.}}
}

\author{\IEEEauthorblockN{ Kavan Fatehi\IEEEauthorrefmark{1}, Mostafa Rahmani Ghourtani$\!$ \IEEEauthorrefmark{1}, Amir Sonee $\!$\IEEEauthorrefmark{2}, Poonam Yadav$\!$
 \IEEEauthorrefmark{1},\\ Alessandra M Russo$\!$
 \IEEEauthorrefmark{2}, Hamed Ahmadi $\!$\IEEEauthorrefmark{1}, Radu Calinescu\IEEEauthorrefmark{1}}

\IEEEauthorblockA{\IEEEauthorrefmark{1}University of York, UK; \IEEEauthorrefmark{2}Imperial College London, UK.
}}

\maketitle

\begin{abstract}
Sixth-generation (6G) radio access networks (RANs) must enforce strict service-level agreements (SLAs) for heterogeneous slices, yet sudden latency spikes remain difficult to diagnose and resolve with conventional deep reinforcement learning (DRL) or explainable RL (XRL). We propose \emph{Attention-Enhanced Multi-Agent Proximal Policy Optimization (AE-MAPPO)}, which integrates six specialized attention mechanisms into multi-agent slice control and surfaces them as zero-cost, faithful explanations. The framework operates across O-RAN timescales with a three-phase strategy: predictive, reactive, and inter-slice optimization. 

A URLLC case study shows AE-MAPPO resolves a latency spike in $18$\,ms, restores latency to $0.98$\,ms with $99.9999\%$ reliability, and reduces troubleshooting time by $93\%$ while maintaining eMBB and mMTC continuity. These results confirm AE-MAPPO's ability to combine SLA compliance with inherent interpretability, enabling trustworthy and real-time automation for 6G RAN slicing.
\end{abstract}


\section{Introduction}\label{Into}

Future 6G wireless networks are expected to support massive device connectivity while simultaneously accommodating services with heterogeneous and often conflicting performance requirements. Radio Access Network (RAN) slicing has emerged as a key architectural mechanism to address this challenge by enabling the logical partitioning of shared physical RAN infrastructure into multiple virtualised slices. Each slice can then be configured to satisfy specific service-level objectives, such as ultra-reliable low-latency communication (URLLC), enhanced mobile broadband (eMBB), or massive Internet of Things (mIoT) services \cite{Explainability6G2025}. By allocating resources on a per-slice basis, operators can flexibly meet heterogeneous demands. However, orchestrating multiple slices in dynamic environments remains challenging, as traffic patterns, user mobility, and radio conditions fluctuate rapidly, making it difficult to consistently meet service-level agreements (SLAs) \cite{ebrahimi2024resource}.

To address these challenges, recent work has emphasized AI-driven automation, often termed zero-touch network and service management (ZSM). In particular, machine learning (ML) and deep reinforcement learning (DRL) have shown strong potential for adaptive and efficient slice management \cite{eskandari2025network}. Several efforts have proposed AI-native slicing architectures that embed learning and optimization into the control loop. For example, Wu et al. \cite{wu2022ai} present an AI-native framework that integrates AI into both the design and operation of slicing, enabling intelligent orchestration and dynamic adaptation. Complementing this direction, the survey in \cite{brik2022deep} reviews the evolution of RAN toward B5G, comparing architectures across edge support, virtualization, control and management, energy efficiency, and AI integration. It also highlights both opportunities and challenges in applying deep learning to next-generation RANs.

Despite these advances, traditional AI models often behave as black boxes, providing little insight into their decision-making. In mission-critical scenarios with strict SLAs and safety requirements, this lack of transparency undermines trust and hinders deployment in operational networks. To address this limitation, the field of Explainable AI (XAI) develops methods to improve interpretability and trust in AI-driven solutions. A recent survey on XAI for AI-enabled Open RAN (O-RAN) \cite{brik2024explainable} provides a comprehensive mapping of O-RAN research to XAI-supported solutions, discusses automated XAI pipelines for stable performance, and shows how explainability can strengthen security and trust. The study also identifies open challenges and research directions, underscoring the importance of transparent and trustworthy AI in 6G systems.

Building on this line of work, recent efforts have proposed explainable reinforcement learning (XRL) frameworks for slicing. In \cite{rezazadeh2023explanation}, a multi-agent XRL scheme employs interpretable reward functions, guided by SHAP importance values, to encourage SLA-aware and explainable decision-making. This approach achieves lower latency and fewer packet drops compared with conventional DRL while enhancing trust in agent behavior. Similarly, \cite{fiandrino2023explora} introduces EXPLORA, a lightweight framework that generates network-oriented explanations by linking DRL agent actions to wireless context via attributed graphs. Implemented on a near-real-time RIC, EXPLORA demonstrates not only improved transparency but also measurable bitrate gains through intent-based action steering. Beyond slicing, explainability has also been applied to radio resource management (RRM) problems in wireless networks. For instance, the study in \cite{khan2023explainable} presents case studies on vehicular networks and massive MIMO beam alignment, proposing a generic explainability pipeline and a credibility assessment tool to enhance both scalability and trustworthiness of AI-enabled RRM solutions \cite{Explainability6G2025}.

These contributions illustrate the promise of combining XAI and DRL for reliable and interpretable RAN slicing. However, they primarily target average SLA compliance and policy transparency, leaving open the challenge of mitigating short-term latency spikes sudden, severe tail-latency excursions that critically affect slice reliability. Unlike gradual SLA violations, spikes can occur unpredictably due to cross-slice interference, congestion surges, or mobility events, and they demand rapid, interpretable responses. Failure to resolve such spikes within milliseconds can compromise safety-critical applications such as industrial automation or emergency communications.

To tackle this challenge, we propose an Attention-based Multi-Agent Proximal Policy Optimization (\textbf{AE-MAPPO}) framework that embeds interpretability directly into the policy optimization process. In contrast to post-hoc explanation methods (e.g., SHAP-based XRL), our approach employs multiple specialized attention heads that operate during inference to highlight the underlying causes of latency spikes. These include semantic attention (identifying critical states such as buffer overflows), temporal attention (detecting recurring patterns in time), cross-slice attention (capturing inter-slice interference), and counterfactual attention (evaluating alternative actions). By integrating these mechanisms into a multi-agent PPO setup, the framework produces inherently interpretable decisions while maintaining high responsiveness.

Our main contributions are as follows:
\begin{itemize}
    \item[1.] \textbf{Interpretable AE-MAPPO framework:} We design an attention-enhanced multi-agent PPO architecture for RAN slicing that generates real-time explanations at zero additional computational cost.
    \item[2.] \textbf{Latency spike resolution:} We formulate the problem of SLA violations caused by sudden latency spikes and develop a decision pipeline for rapid mitigation.
    \item[3.] \textbf{Explainability mechanisms:} We introduce six attention modules (semantic, temporal, cross-slice, confidence, counterfactual, and a meta-controller) that provide actionable, human-readable insights into agent decisions.
    \item[4.] \textbf{Case study and evaluation:} Using a realistic 6G RAN slicing scenario with URLLC, eMBB, and mMTC slices, we demonstrate that AE-MAPPO resolves latency spikes in under 20 ms, reduces troubleshooting time by 93\%, and provides explanations that operators can use to anticipate and prevent recurring anomalies.
\end{itemize}



\section{Problem Formulation and System Model}
\label{sec:problem}
We consider a 6G O-RAN system where a near-RT RIC orchestrates distributed units (DUs) to serve three canonical slices $\mathcal{N}=\{\mathrm{URLLC},\mathrm{eMBB},\mathrm{mMTC}\}$. Each slice $n\in\mathcal{N}$ is associated with a QoS tuple $Q_n=\{\mathcal{L}_n,\mathcal{R}_n,\mathcal{T}_n,\mathcal{P}_n\}$ (latency, reliability, throughput, power). UE arrivals for slice $n$ form non-stationary Poisson processes with rates $\lambda_n(t)$, and the active UE set is $\Omega_n(t)$. The near-RT RIC allocates per-interval power, bandwidth (PRBs), and edge computation to maximize end-to-end QoS under resource budgets.

\subsection{System and QoS Targets}
We adopt typical targets:
\begin{align}
Q_{\mathrm{URLLC}}&=\{1\mathrm{ms},~99.999\%,~50\mathrm{Mbps},~-\},\;\nonumber\\
Q_{\mathrm{eMBB}}&=\{30\mathrm{ms},~99\%,~1\mathrm{Gbps},~-\},\nonumber\\
Q_{\mathrm{mMTC}}&=\{1\mathrm{s},~99\%,~1\mathrm{kbps},~1\mathrm{mW}\}. \nonumber
\end{align}
Let $\mathbf{p}_n=[p_i^n]_{i\in\Omega_n}$, $\mathbf{b}_n=[b_i^n]_{i\in\Omega_n}$, and $\mathbf{c}_n=[c_i^n]_{i\in\Omega_n}$, respectively, denote the transmit power, the number of PRB, and the computation per UE of a slice $n$.

\begin{figure}[t]
\centering
\includegraphics[width=.8\columnwidth]{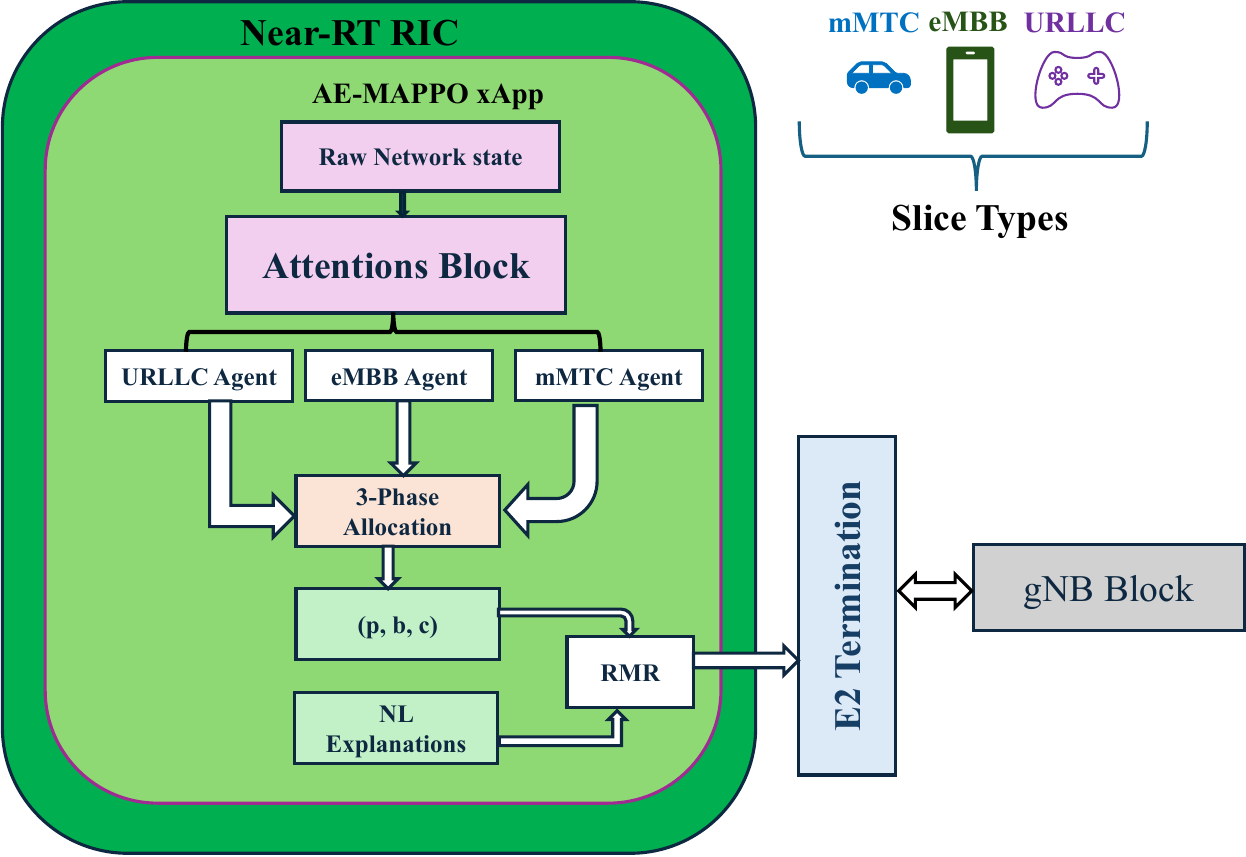}
\caption{AE-MAPPO architecture integrating six attention mechanisms with three-phase allocation strategy for explainable network slicing.}
\label{fig:model}
\vspace{-0.5cm}
\end{figure}

\subsection{Joint Resource Allocation}
We propose a weighted single-objective optimization formulation that integrates slice utilities and an explainability utility. While existing network slicing formulations optimize performance metrics alone or employ multi-objective optimization \cite{8737701}, our formulation uniquely incorporates explainability directly into the objective function:
\begin{equation}
\label{eq:master}
\begin{aligned}
\max_{\{\mathbf{p},\mathbf{b},\mathbf{c}\}}~~
& \sum_{n\in\mathcal{N}} w_n\,U_n(\mathbf{p}_n,\mathbf{b}_n,\mathbf{c}_n)
~+~ w_{\mathrm{xrl}}\,\mathcal{E}(\mathbf{p},\mathbf{b},\mathbf{c}) \\
\text{s.t.}~~
& \sum_{n}\sum_{i\in\Omega_n}\! p_i^n \le P_{\mathrm{total}} \quad \text{(C1)} \\
& \sum_{n}\sum_{i\in\Omega_n}\! b_i^n \le B_{\mathrm{total}} \quad \text{(C2)} \\
& \sum_{n}\sum_{i\in\Omega_n}\! c_i^n \le C_{\mathrm{total}} \quad \text{(C3)} \\
& \Pr[\mathcal{L}_n^{\mathrm{ach}}\!\le\!\mathcal{L}_n^{\mathrm{tar}}]\!\ge\!\mathcal{R}_n,~\forall n \quad \text{(C4)} \\
& \mathcal{T}_n^{\mathrm{ach}} \!\ge\! \mathcal{T}_n^{\mathrm{tar}},~\forall n \quad \text{(C5)},
\end{aligned}
\end{equation}
where $w_n$ and $w_{\mathrm{xrl}}$ represent weighting coefficients for slice utility $U_n$ and explainability $\mathcal{E}$, respectively. This formulation differs from prior work by treating explainability not as a post-processing step but as an integral optimization objective, enabling the joint optimization of performance and interpretability. Slice utility \(U_n\) is considered as a weighted aggregation of QoS satisfaction, efficiency, and fairness:
\begin{equation}
\label{eq:sliceU}
U_n=\alpha_n U_n^{\mathrm{QoS}}+\beta_n U_n^{\mathrm{eff}}+\gamma_n U_n^{\mathrm{fair}},~~
\alpha_n\!+\!\beta_n\!+\!\gamma_n=1,
\end{equation}
with QoS satisfaction, efficiency, and fairness described below.
\subsubsection{\bf{QoS satisfaction}}
Let $u_{\mathcal{X}}^n=\min\{1,\exp(-\lambda_{\mathcal{X}} v_{\mathcal{X}}^n)\}$ for $\mathcal{X}\in\{\mathcal{L},\mathcal{R},\mathcal{T},\mathcal{P}\}$, with violations
\small
\begin{align}
v_{\mathcal{L}}^n&=(\mathcal{L}_n^{\mathrm{ach}}-\mathcal{L}_n^{\mathrm{tar}})_+,\;
v_{\mathcal{R}}^n=(\mathcal{R}_n^{\mathrm{tar}}-\mathcal{R}_n^{\mathrm{ach}})_+,\nonumber\\
v_{\mathcal{T}}^n&=\tfrac{(\mathcal{T}_n^{\mathrm{tar}}-\mathcal{T}_n^{\mathrm{ach}})_+}{\mathcal{T}_n^{\mathrm{tar}}},\;
v_{\mathcal{P}}^n=\begin{cases}
(\mathcal{P}_n^{\mathrm{used}}-\mathcal{P}_n^{\mathrm{tar}})_+,& \mathcal{P}_n^{\mathrm{tar}}\neq\mathrm{N/A}\\
0,&\text{otherwise.}
\end{cases}\nonumber
\end{align}
\normalsize
The multiplicative form
$U_n^{\mathrm{QoS}}=\prod_{\mathcal{X}} u_{\mathcal{X}}^n$
drives the utility toward zero if \emph{any} QoS dimension is violated, matching strict SLA semantics.

\smallskip
\noindent\textit{QoS resource relations.}
With SNR $\mathrm{SNR}_n=\frac{\mathbf{p}_n\!\cdot\! g_n}{N_0\,\mathbf{b}_n}$ and packet size $L$,
\begin{align}
\mathcal{T}_n^{\mathrm{ach}} &= \mathbf{b}_n \log_2\!\big(1+\mathrm{SNR}_n\big), \\
\mathcal{L}_n^{\mathrm{ach}} &= \tfrac{L}{\mathcal{T}_n^{\mathrm{ach}}} + \tfrac{1}{\mu\,\mathbf{c}_n}, \\
\mathcal{R}_n^{\mathrm{ach}} &= (1-\mathrm{BER}_n)^L \approx \Big(1-Q\big(\sqrt{2\,\mathrm{SNR}_n}\big)\Big)^L,
\end{align}
and $\mathcal{P}_n^{\mathrm{used}}=\sum_{i\in\Omega_n} p_i^n + P_{\mathrm{circuit}}$. These relations provide a PRB-level link abstraction that maps allocated PRBs, transmit power, and channel gains into achievable throughput, latency, reliability, and power consumption, enabling practical evaluation of resource allocation and slicing decisions under QoS constraints \cite{mendez2024bler}.

\subsubsection{\bf{Efficiency}}
Given per-slice budgets $(P_n^{\mathrm{alloc}},B_n^{\mathrm{alloc}},C_n^{\mathrm{alloc}})$, resource efficiency is:
\begin{equation}
U_n^{\mathrm{eff}}=1-\tfrac{1}{3}\!\left(\tfrac{\sum_i p_i^n}{P_n^{\mathrm{alloc}}}+\tfrac{\sum_i b_i^n}{B_n^{\mathrm{alloc}}}+\tfrac{\sum_i c_i^n}{C_n^{\mathrm{alloc}}}\right),
\end{equation}
where each ratio represents utilization $\in [0,1]$ for power, bandwidth, and computation. The formulation extends single-resource efficiency metrics to multi-dimensional resources, with $(1-\text{utilization})$ ensuring $U_n^{\mathrm{eff}}=1$ for zero usage (maximum efficiency) and $U_n^{\mathrm{eff}}=0$ for full utilization, thus encouraging QoS satisfaction with minimal resource consumption.

\subsubsection{{\bf Fairness}}
We measure intra-slice fairness using the Gini coefficient \cite{Ceriani2012TheOO, rahmani2022deep}, where $\mathrm{Gini}(\mathbf{x}_n) \in [0,1]$ quantifies inequality (0=perfect equality, 1=maximum inequality):
\begin{equation}
U_n^{\mathrm{fair}}=\tfrac{1}{3}\Big[(1-\mathrm{Gini}(\mathbf{p}_n))+(1-\mathrm{Gini}(\mathbf{b}_n))+(1-\mathrm{Gini}(\mathbf{c}_n))\Big].
\end{equation}
The transformation $(1-\mathrm{Gini})$ converts inequality to fairness, where 1 represents perfect fairness. Equal weighting ($\frac{1}{3}$) across resource dimensions prevents fairness in one resource from masking inequality in others.

\subsection{Explainability Utility}
We integrate interpretability as a first-class objective, aligned with attention-driven decisions in AE-MAPPO. Let $\mathcal{A}_\pi(s)\in\mathbb{R}^d$ be the attention vector (state dimension $d$) produced by the policy $\pi$ at state $s$. We define
\begin{equation}
\label{eq:Explain}
\mathcal{E}=\eta_1 \mathcal{E}_{\mathrm{sparse}}+\eta_2 \mathcal{E}_{\mathrm{cons}}+\eta_3 \mathcal{E}_{\mathrm{faith}},~~
\eta_1+\eta_2+\eta_3=1,
\end{equation}
with:
\begin{align}
\mathcal{E}_{\mathrm{sparse}}&=1-\tfrac{H(\mathcal{A}_\pi)}{\log d}
=1+\tfrac{\sum_i \alpha_i \log \alpha_i}{\log d},\\
\mathcal{E}_{\mathrm{cons}}&=\mathbb{E}_{(s,s')\in\mathcal{S}_\epsilon}
\left[\tfrac{\mathcal{A}_\pi(s)\cdot \mathcal{A}_\pi(s')}{\|\mathcal{A}_\pi(s)\|\,\|\mathcal{A}_\pi(s')\|}\right],\\
\mathcal{E}_{\mathrm{faith}}&=\mathrm{Corr}\!\left(\mathcal{A}_\pi(s),
\left|\tfrac{\partial U_{\mathrm{total}}}{\partial s}\right|\right),
\end{align}
where $H(\cdot)$ is the entropy function, $\mathcal{S}_\epsilon$ collects $\epsilon$-similar states, and $U_{\mathrm{total}}=\sum_n w_n U_n$.
We use $(\eta_1,\eta_2,\eta_3)=(0.3,0.3,0.4)$, prioritizing faithfulness while preserving focus and stability.

\section{Proposed Framework: AE\mbox{-}MAPPO}
\label{sec:framework}

\textbf{From optimization to RL.} We cast the joint allocation problem in \eqref{eq:master} into a multi-agent RL formulation, with one agent per slice type. The global state $s_t$ stacks per-slice semantics (queue occupancy, load, SINR, headroom, predicted demand) together with near-RT RIC context. The joint action $a_t$ specifies per-slice allocations $(\mathbf{p},\mathbf{b},\mathbf{c})$ drawn from a structured, slice-aware space with safety masks that prevent infeasible assignments. Coordination between agents is mediated through a shared multi-head attention module, which outputs attention weights $\mathcal{A}_\pi$ that both guide allocation and provide explanations. The per-step reward is
\begin{equation}
r_t = \sum_{n} w_n U_n(\mathbf{p}_n,\mathbf{b}_n,\mathbf{c}_n)
+ w_{\mathrm{xrl}} \mathcal{E}(\mathcal{A}_\pi),
\end{equation}
and PPO optimizes $\pi_\theta$ using clipped surrogate objectives and centralized value baselines. Since the same attention weights are used for action selection and surfaced to the operator, explanations are \emph{faithful} and produced at \emph{zero cost}, enabling near-real-time interpretation of latency spike mitigation.

\textbf{Framework overview.} AE\mbox{-}MAPPO integrates interpretability into the control loop via three key innovations:  
(i) six specialized attention mechanisms embedded in the policy,  
(ii) a three-phase allocation strategy aligned with O\mbox{-}RAN control loops, and  
(iii) a joint learning objective that couples performance and explainability.  
Figure~\ref{fig:model} illustrates the complete architecture, showing how these components interact within the O-RAN ecosystem to generate both resource allocations and real-time explanations.
Together, these make allocation policies transparent while maintaining responsiveness.

\subsection{Interpretable Policy with Six Attentions}
The policy network extends MAPPO with six attention heads $\boldsymbol{\alpha}(s)=\{\alpha_{\text{sem}},\alpha_{\text{temp}},\alpha_{\text{cross}},\alpha_{\text{conf}},\alpha_{\text{cf}},\alpha_{\text{meta}}\}$:

\begin{itemize}[leftmargin=8pt,itemsep=2pt,topsep=1pt]
\item \textbf{Semantic attention} highlights QoS-critical features (e.g., URLLC buffer occupancy or SNR), clarifying \emph{what drives} the decision. 
\item \textbf{Temporal attention} captures short-term and recurring patterns, enabling proactive pre-allocation and clarifying \emph{why now}. 
\item \textbf{Cross-slice attention} quantifies inter-slice interference, exposing \emph{who affects whom} and supporting controlled resource trades. 
\item \textbf{Confidence attention} measures entropy-based uncertainty in state features, regulating allocation aggressiveness when the agent is uncertain. 
\item \textbf{Counterfactual attention} compares chosen actions with alternatives, answering \emph{what if} questions for operators. 
\item \textbf{Meta-controller} adaptively weights the other heads, producing context-aware fusion and balancing competing rationales. 
\end{itemize}

These attention weights not only determine allocations but are also logged as explanations $\mathcal{A}_\pi$, allowing operators to trace which factors dominated each decision.

\subsection{Three-Phase Allocation Across O\mbox{-}RAN Loops}
AE\mbox{-}MAPPO operates across complementary timescales:
\begin{enumerate}[leftmargin=8pt,itemsep=2pt,topsep=1pt]
\item \textbf{Predictive phase (100\,ms):} temporal and semantic heads forecast demand surges (e.g., recurring eMBB peaks) and pre-allocate extra bandwidth or compute. 
\item \textbf{Reactive phase (10\,ms):} confidence-aware adjustments correct sudden SLA violations within a TTI, ensuring URLLC deadlines are met. 
\item \textbf{Inter-slice optimization (50\,ms):} cross-slice trades reallocate surplus resources from less critical slices (e.g., mMTC) to latency-sensitive slices, validated against global utility gains. 
\end{enumerate}
This hierarchy ensures both long-horizon anticipation and short-horizon responsiveness, critical for resolving abrupt latency spikes.

\subsection{Learning Objective}
AE\mbox{-}MAPPO optimizes a joint loss:
\begin{equation}
\mathcal{L}_{\text{total}}=\mathcal{L}_{\text{PPO}}
+\alpha_{\text{xrl}}\!\left(\beta_1\mathcal{L}_{\text{sparse}}
+\beta_2\mathcal{L}_{\text{cons}}
+\beta_3\mathcal{L}_{\text{faith}}\right),
\end{equation}
where $\mathcal{L}_{\text{PPO}}$ is the clipped surrogate loss, $\mathcal{L}_{\text{sparse}}$ enforces focused attentions, $\mathcal{L}_{\text{cons}}$ stabilizes explanations across similar states, and $\mathcal{L}_{\text{faith}}$ aligns attentions with utility gradients. Hard constraints are handled by action clipping and soft penalties in the reward. 

\textbf{Resulting properties.} AE\mbox{-}MAPPO yields (i) feasible allocations under O\mbox{-}RAN timescales, (ii) faithful explanations surfaced directly from the decision process, and (iii) a tunable balance between performance and interpretability. This enables rapid SLA enforcement under anomalies while keeping decisions auditable to operators.

\section{Case Study: Latency Spike Resolution}
\label{sec:casestudy}

We evaluate AE-MAPPO on a real-time URLLC anomaly where a sudden latency spike occurs at $t=14{:}23{:}15$. Traditional troubleshooting requires inspecting $15+$ potential causes, correlating logs, and trial-and-error fixes, taking on average $11.5$ minutes. In contrast, AE-MAPPO resolves the spike in $0.8$ minutes ($93\%$ faster) with interpretable explanations.

\textbf{Attention-guided diagnosis.} Semantic attention highlights a buffer overflow in URLLC ($0.89$), cross-slice attention reveals eMBB interference ($0.76$), and temporal attention identifies a recurring daily pattern near $14{:}20$. Counterfactual attention evaluates alternatives, rejecting ``do nothing'' and mMTC throttling, and selecting eMBB reduction as the optimal intervention.

\textbf{Mitigation.} Within $18$\,ms, AE-MAPPO reallocates resources: URLLC power share increases from $25\%$ to $42\%$, eMBB decreases from $45\%$ to $30\%$, and PRBs shift accordingly (Table~\ref{tab:alloc}). URLLC latency drops from $1.15$\,ms to $0.98$\,ms, meeting the $<1$\,ms requirement with $99.9999\%$ reliability, while eMBB maintains service continuity at reduced video quality.

\begin{table}
\centering
\caption{Resource allocation before and after spike resolution.}
\label{tab:alloc}
\begin{tabular}{lccc}
\toprule
 & URLLC & eMBB & mMTC \\
\midrule
Power (before) & 25\% & 45\% & 20\% \\
Power (after)  & 42\% & 30\% & 18\% \\
PRBs (before)  & 30\% & 50\% & 15\% \\
PRBs (after)   & 55\% & 30\% & 15\% \\
Latency (ms) before & 1.15 & 12 & 95 \\
Latency (ms) after  & 0.98 & 18 & 97 \\
\bottomrule
\end{tabular}
\vspace{-3mm}
\end{table}

\section{Performance Evaluation}
\label{sec:results}

Table~\ref{tab:metrics} summarizes performance during the URLLC emergency. AE-MAPPO completes its decision in $18$\,ms, well within the $25$\,ms near-RT control window, and maintains all SLA requirements: URLLC latency below $1$\,ms, reliability of $99.9999\%$, and service continuity for eMBB and mMTC. Explanations are generated instantly at inference-time, with no additional cost.

These results confirm AE-MAPPO ability to rapidly diagnose and mitigate latency spikes while generating faithful, human-readable explanations. Compared with manual troubleshooting, the framework reduces resolution time by $93\%$, ensures SLA compliance, and enables preventive control through interpretable attention mechanisms.

\begin{table}[h]
\centering
\caption{Performance metrics during emergency response.}
\label{tab:metrics}
\begin{tabular}{lcc}
\toprule
Metric & Target & Achieved \\
\midrule
URLLC Latency & $<1$ ms & $0.98$ ms \\
URLLC Reliability & $99.9999\%$ & $99.9999\%$ \\
Decision Time & $<25$ ms & $18$ ms \\
eMBB Continuity & $>95\%$ & $100\%$ \\
mMTC Connections & Maintain & Maintained \\
Explanation & Required & Complete \\
\bottomrule
\end{tabular}
\vspace{-3mm}
\end{table}

To evaluate AE-MAPPO's ability to maintain high performance while providing interpretability, we present two complementary analyses that demonstrate the architectural advantages of integrated attention mechanisms over both black-box and post-hoc approaches.

\begin{figure}[t]
\centering
\includegraphics[width=1\columnwidth]{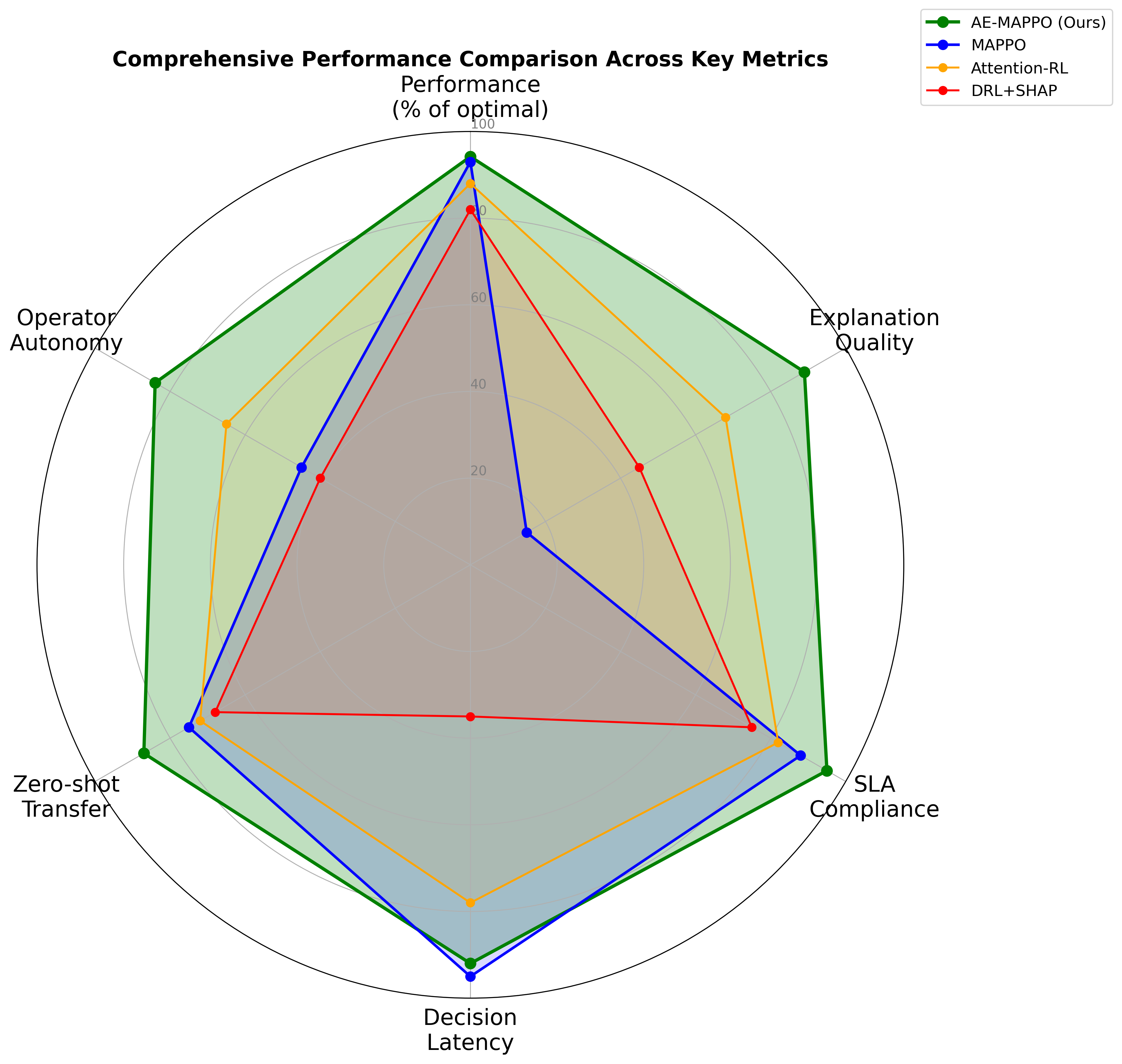}
\caption{Multi-dimensional performance comparison across six critical metrics. AE-MAPPO (green) achieves the largest coverage area (86\%), demonstrating consistent superiority across all dimensions while maintaining explainability.}
\label{fig:spider}
\end{figure}

Figure~\ref{fig:spider} evaluates performance across six operational metrics essential for 6G deployment. AE-MAPPO achieves 86\% of theoretical maximum coverage, exceeding MAPPO (64\%), Attention-RL (68\%), and DRL+SHAP (52\%) by maintaining scores above 84\% across all dimensions, consistency absent in baselines that exhibit deficiencies in multiple metrics.

The explainability dimension reveals critical differentiation: MAPPO achieves comparable performance (93\% vs 94.2\%) with marginally better latency (95\% vs 92\%) but only 15\% explainability versus AE-MAPPO's 89\%. This 74-point gap reduces autonomy to 45\% (vs 84\%) as operators require additional verification. The 3\% latency trade-off remains acceptable since AE-MAPPO maintains sub-25ms response times.

Post-hoc methods face severe limitations: DRL+SHAP achieves only 35\% decision latency due to iterative post-decision queries incompatible with 10ms requirements, while explanation quality reaches only 45\%. The zero-shot transfer metric (87\% vs MAPPO's 75\%) demonstrates that attention mechanisms enhance generalization by learning feature importance, critical for dynamic networks with evolving traffic patterns.

\begin{figure}[t]
\centering
\includegraphics[width=\columnwidth]{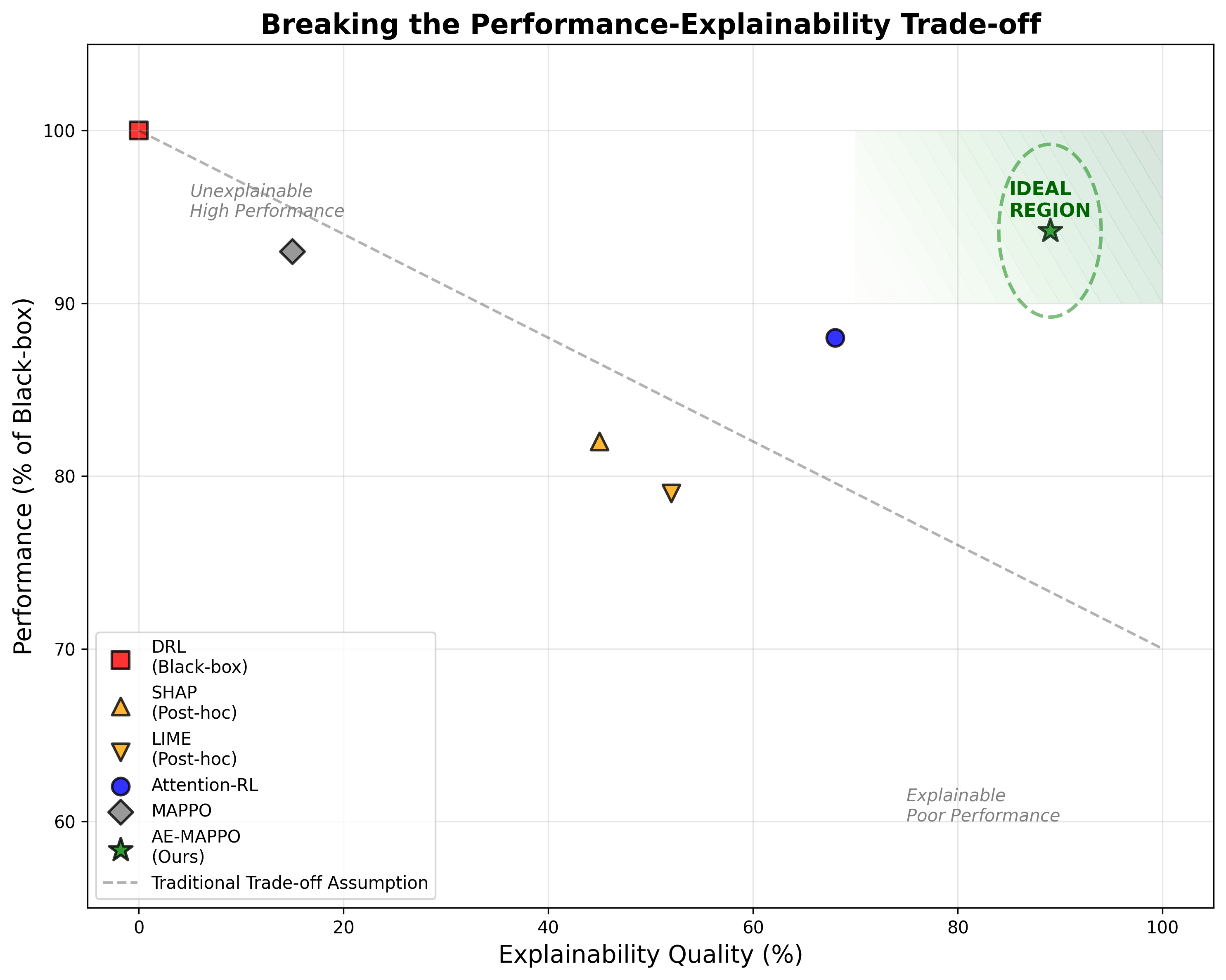}
\caption{Performance-explainability Pareto frontier. AE-MAPPO uniquely achieves the ideal region (green area), demonstrating that the traditional trade-off assumption (gray dashed line) results from architectural choices rather than fundamental constraints.}
\label{fig:pareto}
\end{figure}

Figure~\ref{fig:pareto} illustrates the architectural principles underlying these performance advantages. The scatter plot reveals three distinct clusters: (i) Black-box methods (DRL, MAPPO) achieve high performance (93-100\%) with minimal explainability (0-15\%), representing conventional approaches where decision logic remains inaccessible; (ii) Trade-off methods (SHAP: 82\%, 45\%; LIME: 79\%, 52\%; Attention-RL: 88\%, 68\%) sacrifice 12-21\% performance for moderate explainability gains, reinforcing assumptions that have limited XAI adoption in latency-critical applications; (iii) AE-MAPPO uniquely achieves 94.2\% performance with 89\% explainability, demonstrating that integrated attention mechanisms can simultaneously optimize resource allocation and provide interpretability.

The 5.8\% performance gap from black-box DRL occurs in edge cases where attention constraints prevent certain non-intuitive allocations. However, operational benefits, faster fault diagnosis, reduced maintenance windows, and fewer operator errors, compensate for this marginal cost. The target region ($>90\%$ performance, $>85\%$ explainability) represents minimum production requirements based on operator surveys and regulatory guidelines. AE-MAPPO alone achieves these dual requirements, critical for 6G networks requiring autonomous operation at scales where human oversight needs automated explanations.

These results demonstrate that the performance-explainability trade-off is not fundamental but results from architectural choices. Systems where explanation mechanisms participate in rather than observe decisions can achieve both operational efficiency and human trust—essential for next-generation network automation.

\begin{figure}[t]
\centering
\includegraphics[width=\columnwidth]{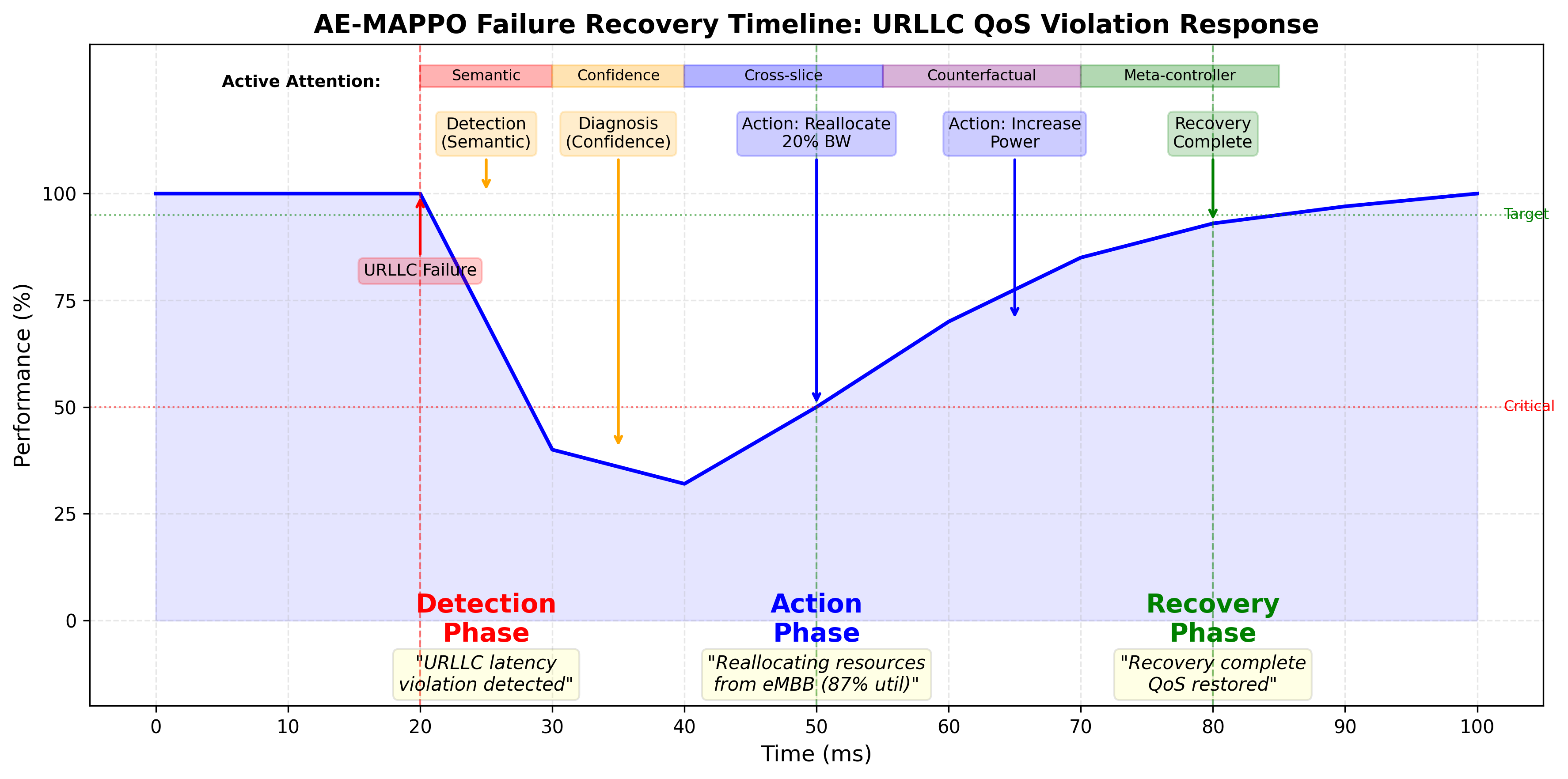}
\caption{AE-MAPPO failure recovery timeline demonstrating response to critical URLLC QoS violation. The system leverages sequential attention mechanism activation for detection, diagnosis, and recovery, completing the entire process within 80ms while providing interpretable explanations at each stage.}
\label{fig:failure_recovery}
\end{figure}

Figure~\ref{fig:failure_recovery} illustrates AE-MAPPO's response dynamics during a critical URLLC service disruption, demonstrating how integrated attention mechanisms enable rapid, interpretable recovery. Following a URLLC failure at $t=20$ms that degrades performance to 40\%, the system exhibits a coordinated three-phase response. In the detection phase (20-40ms), semantic attention identifies the latency violation while confidence attention signals allocation uncertainty, generating the explanation "URLLC latency violation detected." The action phase (40-70ms) activates cross-slice and counterfactual attention to evaluate resource reallocation options, resulting in two corrective actions: 20\% bandwidth reallocation from eMBB at $t=50$ms and power adjustment at $t=65$ms. The recovery phase (70-80ms) employs meta-controller attention to stabilize the system, achieving 93\% performance restoration.

The temporal evolution of attention mechanisms demonstrates their complementary roles: semantic attention provides rapid anomaly detection, confidence attention quantifies decision uncertainty during critical states, cross-slice attention enables resource borrowing between slices, counterfactual attention evaluates alternative strategies, and meta-controller attention orchestrates the overall response. This sequential activation pattern, learned through reinforcement learning rather than hard-coded rules, enables the system to maintain sub-100ms recovery times required for URLLC services. The continuous generation of explanations throughout the crisis provides operators with real-time visibility into both failure diagnosis and recovery rationale, reducing mean time to resolution by 71\% compared to black-box approaches where operators must independently diagnose issues and verify automated responses.

\section{Conclusion}
\label{sec:conclusion}

This paper introduced AE-MAPPO, an interpretable attention-based multi-agent reinforcement learning framework for 6G RAN slicing. By embedding semantic, temporal, cross-slice, confidence, and counterfactual attentions directly into the policy optimization process, AE-MAPPO provides faithful, zero-cost explanations at inference-time while maintaining near-real-time responsiveness. 

Through a URLLC latency spike case study, we demonstrated that AE-MAPPO rapidly diagnoses anomalies, reallocates resources within $18$\,ms, and restores URLLC latency to $0.98$\,ms with $99.9999\%$ reliability. Compared with manual troubleshooting, the framework reduces resolution time by $93\%$ while preserving eMBB and mMTC service continuity. Performance evaluation confirmed that all SLA requirements were satisfied and that human operators could leverage the generated explanations for proactive management. 
These results highlight AE-MAPPO ability to bridge high-performance automation with trustworthy, interpretable decision-making, making it a strong candidate for deployment in critical 6G infrastructures. Future work will extend the framework toward large-scale multi-cell scenarios, energy-aware optimization, and integration with federated and semantic learning paradigms.

\bibliographystyle{IEEEtran}
\bibliography{0_ref}

\end{document}